\documentclass[final,3p,times,twocolumn,sort&compress]{elsarticle}
\usepackage{graphicx}
\usepackage{float}
\usepackage{color}
\usepackage{amsmath}

\begin{document}
\begin{frontmatter}
\title{Influence of cylindrical geometry and alignment layers on growth process and selective reflection of blue phase domains} 
\date{\today}
\author{M. M. Sala-Tefelska\corref{cor1}\fnref{label1}}
\ead{martef@if.pw.edu.pl}
\author[label1]{K. Orzechowski}
\author[label1]{M. Sierakowski}
\author[label1]{A. Siarkowska}
\author[label1]{T. Woli\'nski}
\author[label2]{O. Strze\.zysz}
\author[label2]{P. Kula}
\address[label1]{Warsaw University of Technology, Faculty of Physics, 00-662 Warsaw, Poland}
\address[label2]{Military University of Technology, Institute of Chemistry, 00-908 Warsaw, Poland}
\begin{abstract}
In this work the influence of cylindrical shape and alignment layers on light reflection in Blue Phase Liquid Crystal (BPLC) is presented. For the first time, the process of BP domains growth in a capillary is presented. The cylindrical structure, its diameter and alignment layers change the orientations of cubic blue phase (BP) domains and affect their growth. 
By using temperature and external electric field the uniform structure was obtained. In this study the ability of switching between BP\,I and chiral phase in a capillary is also shown.
\end{abstract}
\end{frontmatter}
%\maketitle  
\section{Introduction}
\begin{figure}
\centering
\includegraphics[width=\columnwidth]{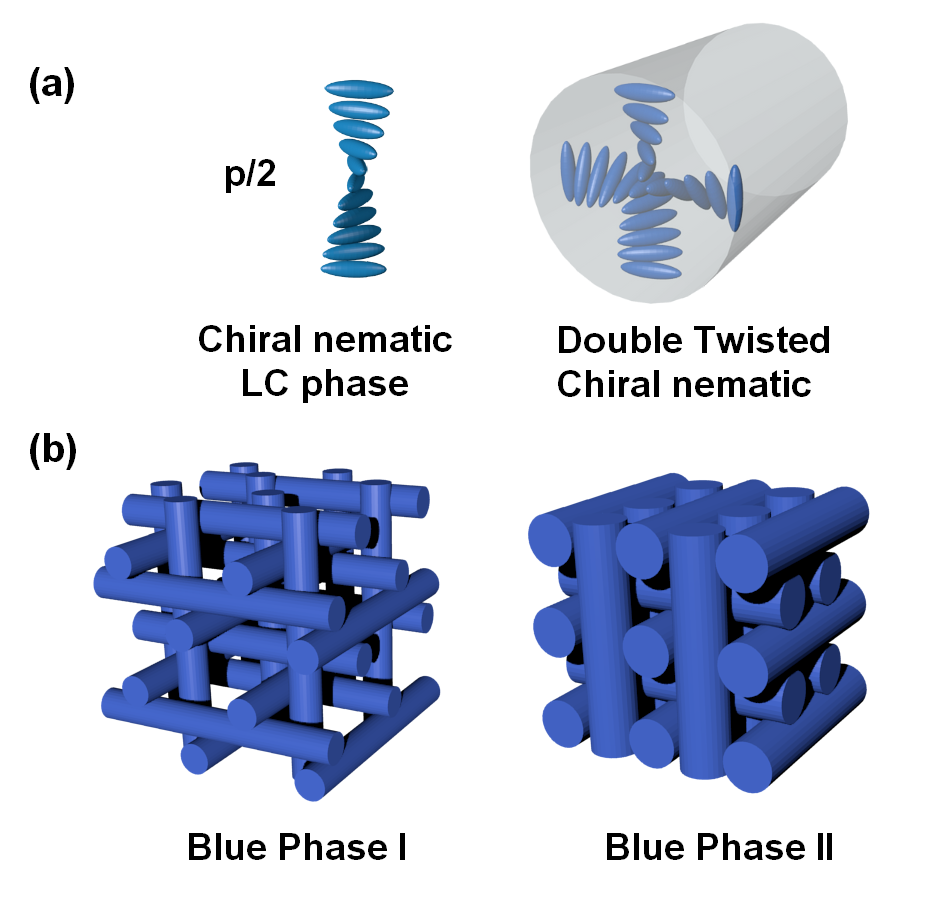}
\caption{(a) Single helical structure (half of pitch $p$) of the chiral nematic LC (left) and one of the BP cube components - cylinder with double twisted chiral nematic (right) (b) Cubic structures of Blue Phase I and II.}
\label{fig:BP}
\end{figure}

The blue phase (BP) is a type of a liquid crystal state between isotropic and chiral phase. It occurs in a narrow range of temperatures in chiral nematic material with a small helix pitch. The BP has a double twisted cylinders (Fig.\ \ref{fig:BP}a) which are placed on the cubic lattice in a various directions. The double twisted cylinders are separated by the network of disclination lines which stabilize 3-dimensional periodic structure of BP. This is a metastable state with minimal energy, which explains the very narrow temperature range of occurrence of BP phase. The double twisted chiral nematic LC in BP phase resembles chiral nematic phase in capillaries (Fig.\ \ref{fig:BP}a) \cite{Kitzerow96, MTefelska}. The BP can exist in three distinct sub-phases: blue phase I (BP\,I), II (BP\,II) and III (BP\,III) also known as a fog phase. The BP\,I and BP\,II (Fig.\ \ref{fig:BP}b) have a cubic symmetry and exhibit the Bragg reflections. The frequencies of reflected light are in visible and UV ranges. The BP\,III has the same symmetry as the isotropic phase and a foggy, uniform consistency \cite{Kitzerow01, Henrich2011}. 
Despite the blue phase was discovered already in 1888, by Austrian botanist Reinitzer \cite{Reinitzer1888}, its properties were described years later at the end of 20th century. Currently the new technology of observation and analysis allow to get more information about structure of BP domains, their growth \cite{PieranskiBP_growth, PieranskiBP_growth2, Chen14_elect_growth}, optical properties and potential applications \cite{BLUMEL1984, ONUSSEIT1983, Chen12Hysteresis, Lin2011Measuring, Lin11, Chen16, Tanaka15, Castles_BP_applications, Orzechowski2017_1, Orzechowski2017_2}. The temperature range of BP appearance can be extended, by addition of bimesogenic molecules or nanoparticles\cite{Huang_WideBPrange_mesogens,Coles05_BPwidetemp,Wanli_wide_BP_various_method} and stabilized by using a polymer \cite{Kikuchi02}. Therefore, it is easy to observe and use BP phase in many configurations.   
In the paper \cite{Poudereux14} the thermal tuning of the guided light in photonic crystal fiber filled with BPLC was observed. 
M. Wahle et al. \cite{Kitzerow16} shown experimental and modal analyses of electric field applied on photonic crystal fiber filled with BPLC. They observed that asymmetric shift of the photonic band gaps is a result of the mixed polarization. The BP does not need orientation layers which is a huge advantage for liquid crystal technology. In spite of this, the layers can force a certain arrangement of the BP cubes, that can be used for other purposes. 
Recently it was noticed that alignment layers change the Bragg reflection in BP cells \cite{Pankaj15, Chen16, Orzechowski2017_1}. It is not obvious how BP behaves in different geometries and how orientation layer influence on the liquid-crystal configuration. Continuing the research in this field we decided to examine cholesteric LC with BP phase in micro-capillaries. 
In this work, the BP domains growth, in a single cylinder structure, was observed. Also the influence of alignment layers and electric field on selective Bragg reflection was presented.   
\section{Materials and setup}
\begin{figure}
\includegraphics[width=\columnwidth]{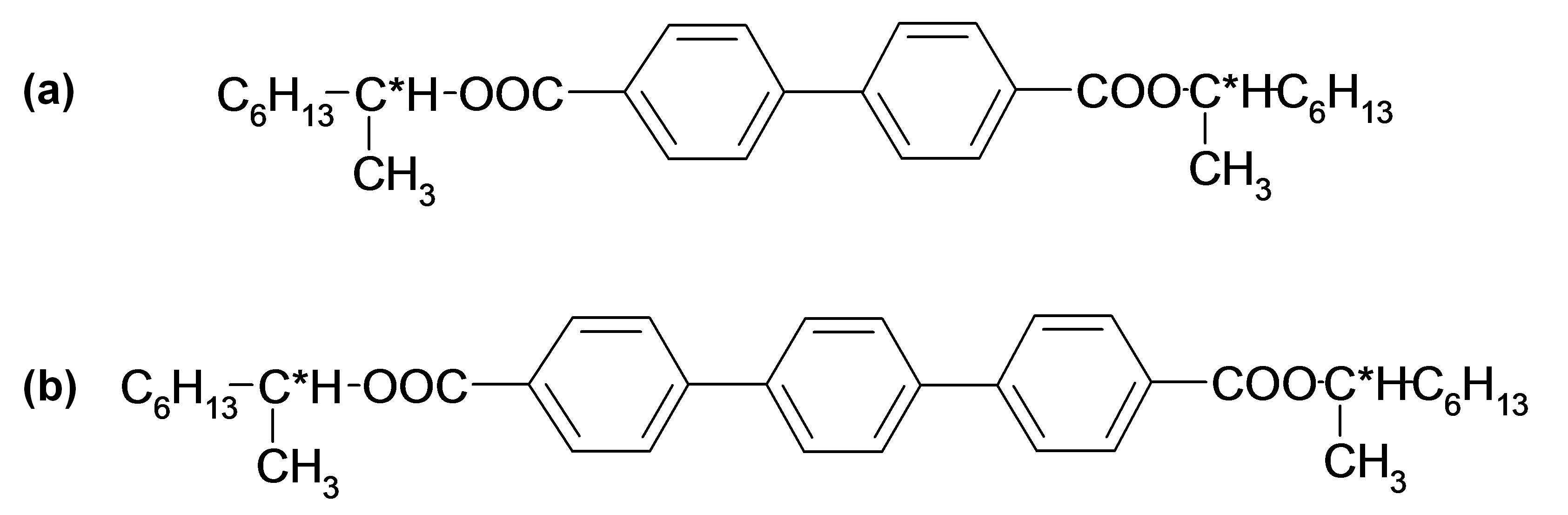}
\caption{(a) OAD1 - biphenyl-4,4’-dicarboxylic acid bis-(1-methylheptyl) ester (b) OAD2 - [1,1’;4’,1”]terphenyl-4,4”-dicarboxylic acid bis-(1-methylheptyl) ester} \label{fig:formulas}
\end{figure}

The BPLC material used in the experiment is composed of nematic LC host - 1912 mixture mixed with two chiral dopants \cite{Kula_13}. The major compositions of 1912 mixture are photochemically stable fluorinated oligophenyls with fluorinated cyclohexyl- and bicyclohexylbiphenyls \cite{Chojnowska_14}. The compound structures and concentrations are listed in Table 1. Compounds 4-10 were synthesized at the Military University of Technology. Compounds 1-3 and 11 were provided by Valiant Fine Chemicals. The 1912 mixture exhibits melting point below $-20^{\circ}\, \text{C}$, clearing point at $78^{\circ}\, \text{C}$ and has medium birefringence 
$\Delta n=0.178$ at 589 nm. It has relatively low electric anisotropy $\Delta \varepsilon=12.6$ at 1kHz, and medium rotational viscosity $\gamma =305\; \text{mPa}\cdot \text{s}$. All parameters were measured at $20^{\circ}\, \text{C}$. To obtain chiral nematic mixture with BP phase, two optical active dopants (OADs) were added to 1912 mixture: biphenyl-4,4’-dicarboxylic acid bis-(1-methylheptyl) ester and [1,1’;4’,1”]terphenyl-4,4”-dicarboxylic acid bis-(1-methylheptyl) ester, both synthesized at the Military University of Technology \cite{Kula_13}. The both structural formulas are presented in Fig.\ \ref{fig:formulas}. The concentration of each optically active compound was 7.0 wt\%. The macroscopic helical twisting power (HTP) of biphenyl and terphenyl ester measured at $20^{\circ}\, \text{C}$ in mixture of fluorinated compounds analogous to components of mixture 1912 was 25 $\mu \text{m}^{-1}$  and  30 $\mu \text{m}^{-1}$, respectively. The helical pitch was $571\,\text{nm}$ and  $476\,\text{nm}$ respectively to the HTP. In this composition the BP\,II appears in a range of temperatures from $60.5^{\circ}\, \text{C}$ to $59^{\circ}\, \text{C}$ and BP\,I appears from $58.9^{\circ}\, \text{C}$ to $54^{\circ}\, \text{C}$, in a cooling process. 
%\vspace{10pt} 
\begin{table}
\caption{\label{Table 1} Composition of nematic mixture 1912}
\begin{tabular}{|p{0.35cm}|p{4cm}|p{1.7cm}|}
\hline
	 No. & \centering{Formula} &\footnotesize {Concentration  in weight [\%]} \\ 
\hline
\hline
	 1 &  \includegraphics[scale=0.5]{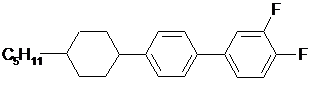} & \centerline {19.8}  \\ 
\hline
	 2 & \includegraphics[scale=0.5]{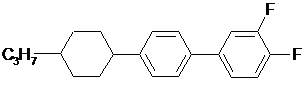} & \centerline {7.2} \\ 
\hline
	 3 & \includegraphics[scale=0.5]{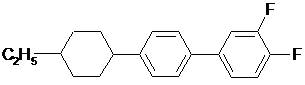} & \centerline {6.5} \\ 
\hline
	 4 & \includegraphics[scale=0.4]{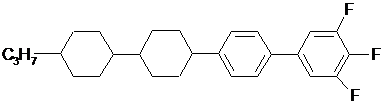} & \centerline {10.8} \\ 
\hline
	 5 & \includegraphics[scale=0.4]{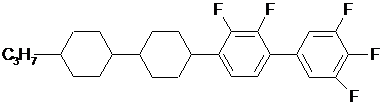} & \centerline {2.5} \\ 
\hline
	 6 & \includegraphics[scale=0.5]{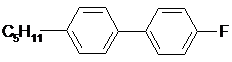} & \centerline {12.5} \\
\hline
	 7 & \includegraphics[scale=0.5]{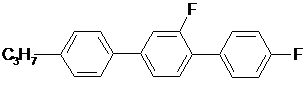} & \centerline {7.1} \\
\hline
	 8 & \includegraphics[scale=0.5]{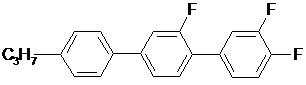} & \centerline {6.1} \\
\hline
	 9 & \includegraphics[scale=0.5]{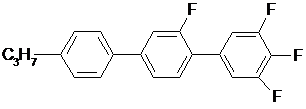} & \centerline {12.9} \\
\hline
	 10 & \includegraphics[scale=0.5]{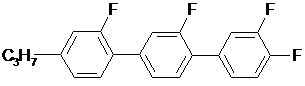} & \centerline {8.4} \\ 
\hline
	 11 & \includegraphics[scale=0.4]{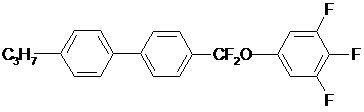} & \centerline {6.3} \\ 
\hline
\end{tabular}
\end{table}

To observe BP domains in the capillaries the Nikon Eclipse Ts2R-FL polarized light microscope and Linkam THMS600 heating stage were used. The capillary was made of pure silica glass and had 60 $\mu \text{m}$ of inner diameter with about 200 $\mu \text{m}$ of outer diameter. Cylinders were filled with the chiral nematic mixture by capillary action. Capillary with BPLC was put into the heating chamber and placed in the polarizing microscope. The inner surface of capillaries was covered by SE130 and SE1211 polyimides (produced by Nissan Chemical Industries, Ltd.) to obtain planar and homeotropic alignment layer respectively. 
\section{Experiments and discussions}
\begin{figure}
\includegraphics[width=\columnwidth]{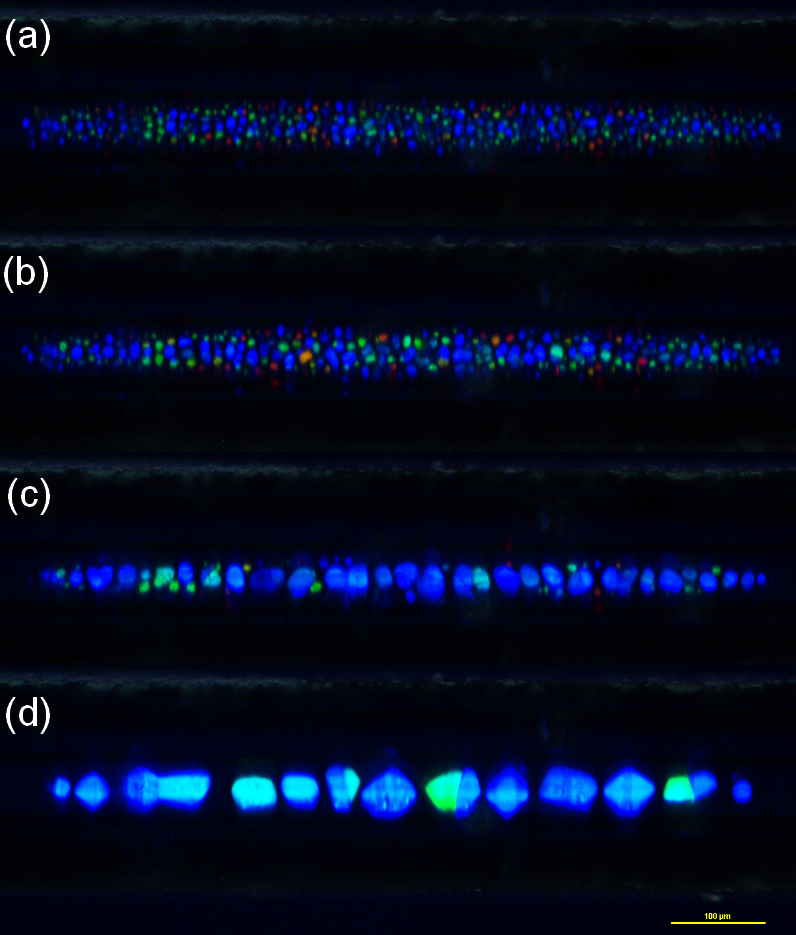}
\caption{Growth of BP\,II domains, after: (a) 0 min. (b) 2 min. (c) 11 min. (d) 55 min.\ left at constant temperature of $60.5^{\circ}\, \text{C}$. Capillary without any alignment layers.}
\label{fig:growth}
\end{figure}

In order to investigate the growth of BP crystal in the capillary, heating and cooling processes were used. First of all the BP capillary, without any alignment layer (NAL), with 60 $\mu \text{m}$ in diameter, was heated to obtain the isotropic phase $70^{\circ}\, \text{C}$. Next, during a slow cooling process, with step $0.1^{\circ}\, \text{C} \text{/min}$ the small domains of BP\,II were observed in a capillary (Fig.\ \ref{fig:growth}a). At that moment the cooling was stopped and the BP capillary was left at stabilized temperature of $60.5^{\circ}\, \text{C}$ for about an hour. The growth of BP domains in a capillary is presented in Fig.\ \ref{fig:growth}. As can be seen, the crystal forms grow upto the inner dimension of the capillary and each large domain can be distinguished. The obtained BP\,II domains are separated and are uniformly distributed in the capillary. Their shape sometimes resemble natural diamond crystals (Fig.\ \ref{fig:growth}d). The cylindrical structure has different anchoring conditions. It causes that the greater part of the BP domains, located in the central part of capillary, reflect the blue light. On the border of capillary, occasionally, red and green reflections appear. 
\begin{figure}
\includegraphics[width=\columnwidth]{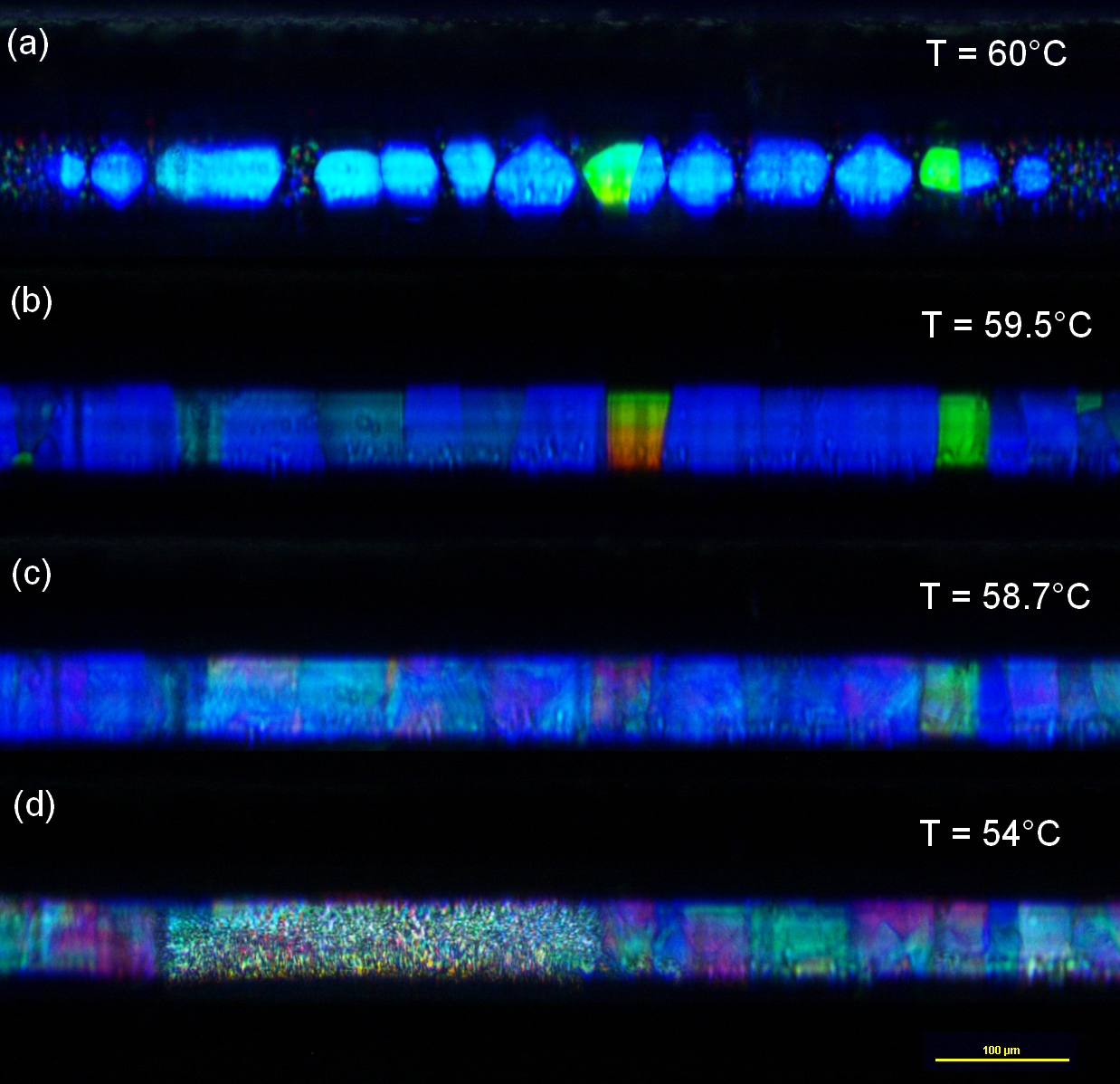}
\caption{(a)-(c) Capillary filled with BPLC during slow cooling process to obtain a quasi-uniform structure, (a),(b) BP\,II, (c) BP\,I, (d) transition from BP\,I to the chiral phase. \label{fig:homogeneity}}
\end{figure} 
When the large domains were obtained, the slow cooling process was applied (Fig.\ \ref{fig:homogeneity}a-c). Then, the quasi-uniform structure of BP\,II (Fig.\ \ref{fig:homogeneity}b) and then BP\,I in capillary was obtained (Fig.\ \ref{fig:homogeneity}c). The selective reflection of light occurs in a blue and pink spectral ranges. The BP\,I transition to the chiral phase was observed at $54^{\circ}\, \text{C}$ (Fig.\ \ref{fig:homogeneity}d).  

\begin{figure}
\includegraphics[width=\columnwidth]{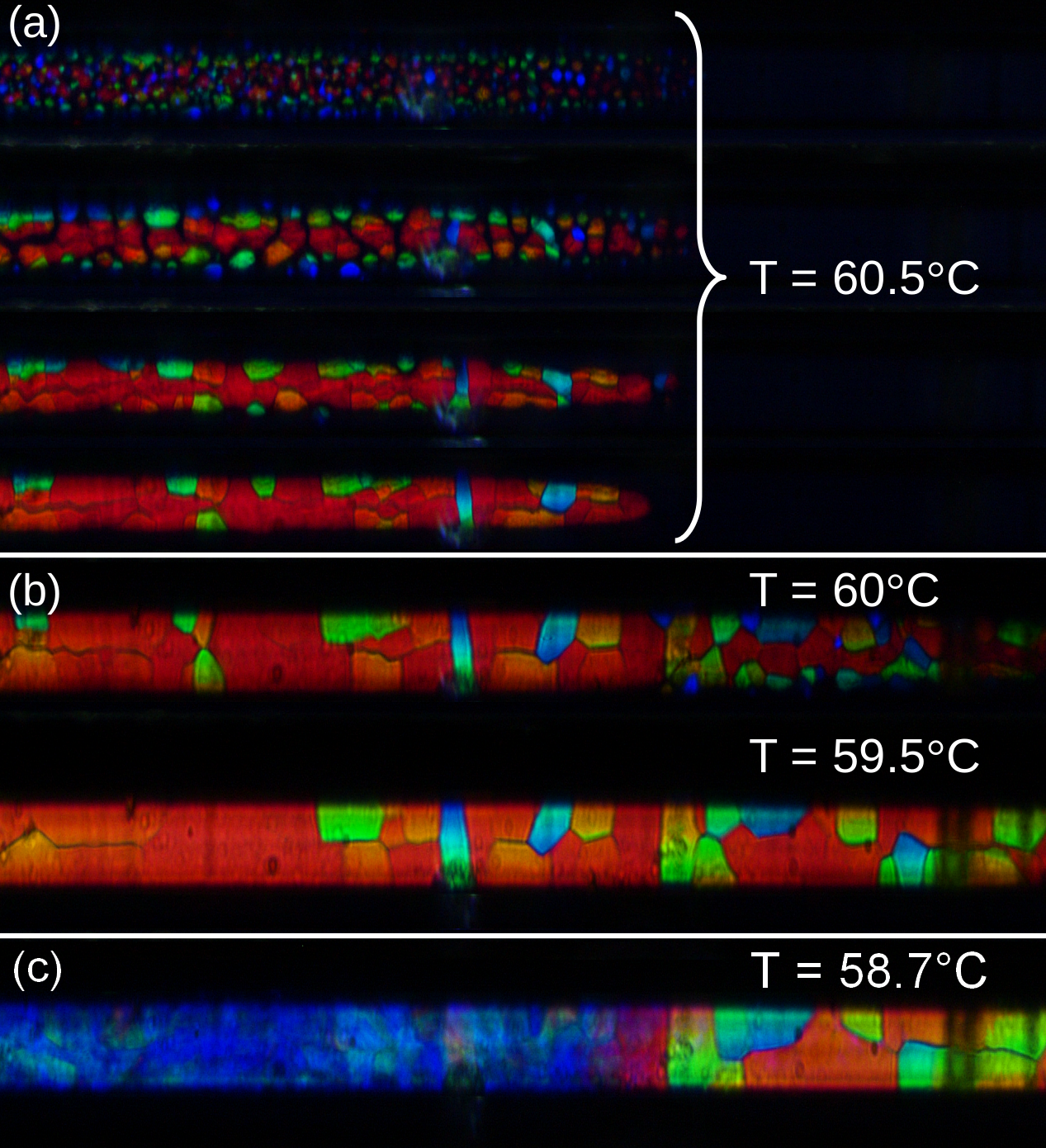}
\caption{Capillary with HA filled with BP LC: (a) the growth of BPII domains, (b) a quasi-uniform structure of BP\,II, (c) transition from BP\,II to BP\,I.}
\label{fig:HG}
\end{figure}

\begin{figure}
\centering
\includegraphics[width=.47\textwidth]{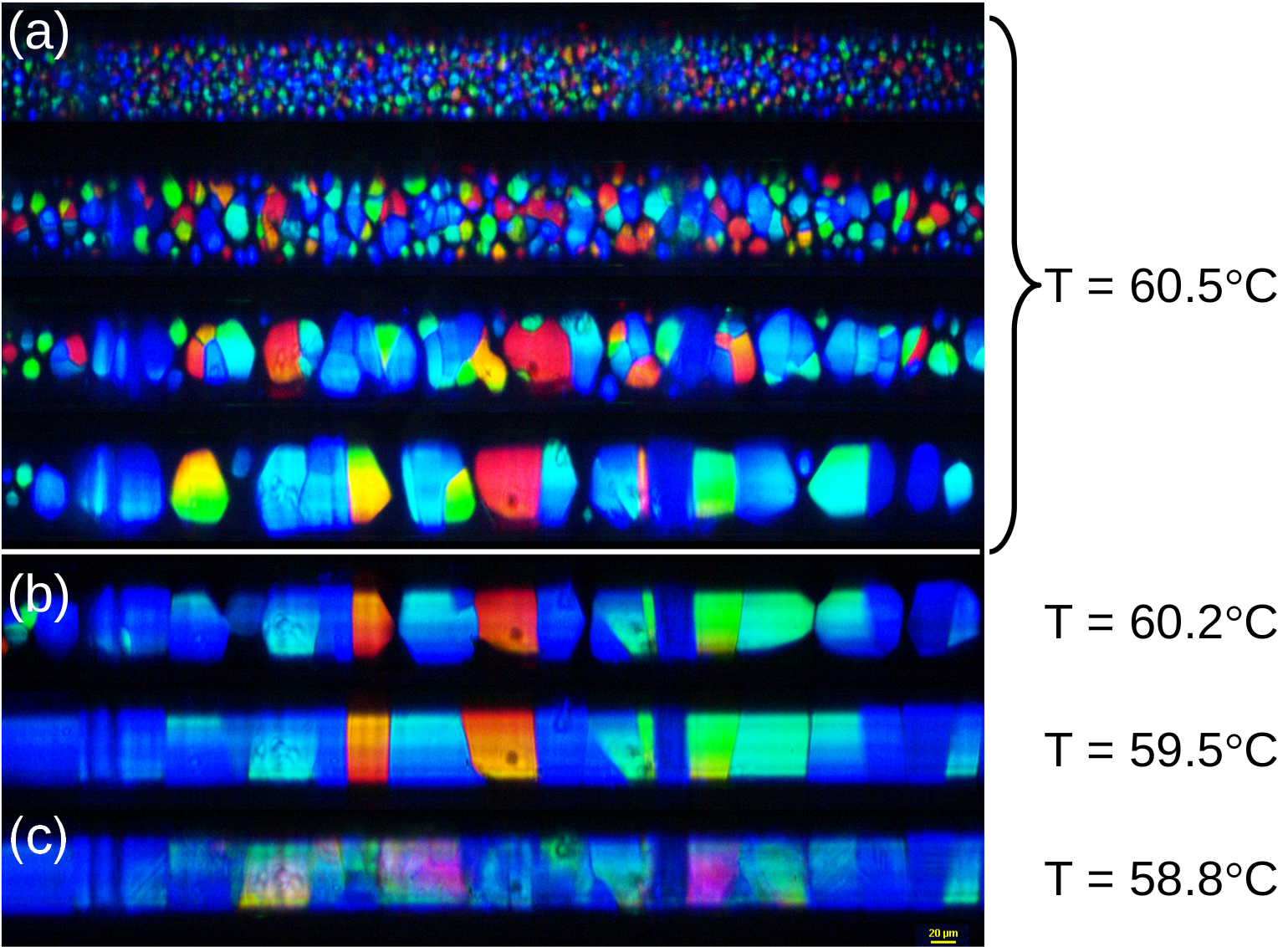}
\caption{Capillary with VA filled with BP LC: (a) the growth of BPII domains, (b) a quasi-uniform structure of BP\,II, (c) BP\,I.}
\label{fig:HT}
\end{figure}

\begin{figure}
\includegraphics[width=\columnwidth]{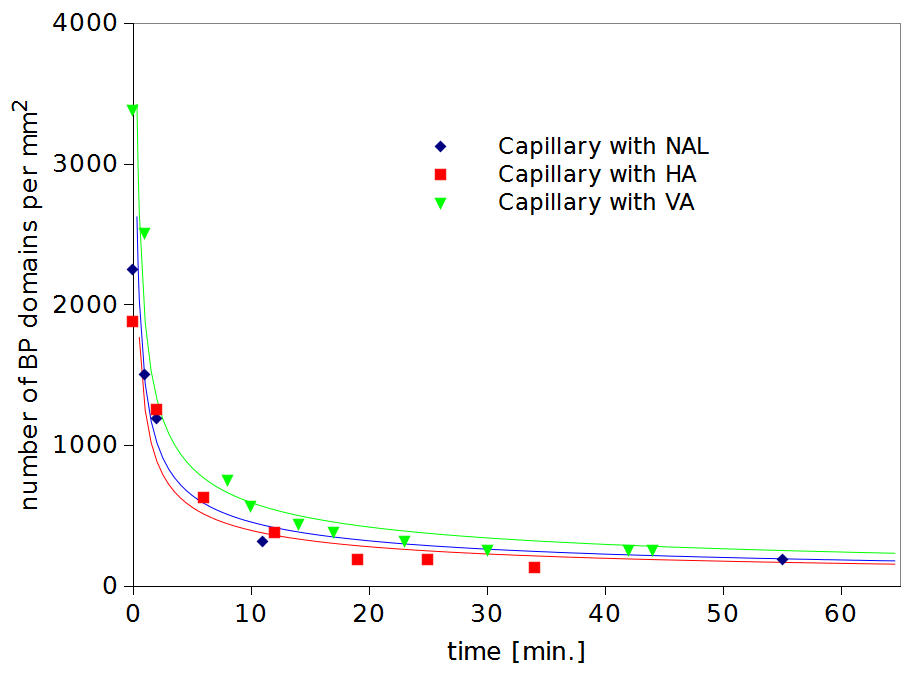}
\caption{Number of BP domains per mm${^2}$ in a capillary versus time.} 
\label{fig:number}
\end{figure}

Next, the inner surface of capillary was covered with homogenous or homeotropic alignment layer causing horizontal (HA) or vertical (VA) anchoring, respectively. Capillaries with HA and VA anchoring layer were filled with BPLC. To obtain BP domains growth, the same stabilization process, previously described, has been applied. For HA, the BP\,II domains reflect red light. For the capillary with HA the small BP domains, on the edges of the cylinder, mostly reflect green light, while the central space reflects red light. In the process of growth, BP domains connect together creating a quasi homogeneous structure (Fig.\ \ref{fig:HG}a), otherwise than in a capillary with NAL and VA. 
By additional, slow cooling process to $59.5^{\circ}\, \text{C}$ it was possible to get uniform structure reflecting red light (Fig.\ \ref{fig:HG}b). For $58.7^{\circ}\, \text{C}$ the phase transition from BP\,II to BP\,I was observed (Fig.\ \ref{fig:HG}c). For the VA, in the same temperature, much more small domains with different selective reflection wavelengths were observed (Fig.\ \ref{fig:HT}a). Slow cooling process allowed to obtain single large domains (Fig.\ \ref{fig:HT}b). Also the uniform structure with predominance of blue light reflection was obtained. The average number of BP domains per square millimeter as a function of time is calculated and is shown in Fig.\ \ref{fig:number}. 
It was determined that the average number of BP domains is nearly proportional to the reciprocal of the square root of time (see fitted lines Fig.\ \ref{fig:number}). At the first stage the process of BP domains growth is rapid. For the VA a lot of BP domains are created, more than for HA or NAL.     

\begin{figure}
\includegraphics[width=\columnwidth]{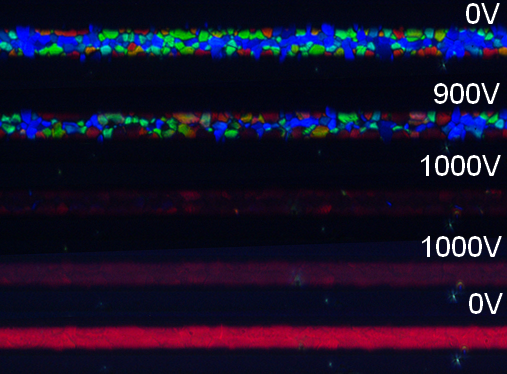}
\caption{BPLC capillary with BP\,II at $58^{\circ}\, \text{C}$ under the influence of electric field.} 
\label{fig:EF_BP}
\end{figure}

\begin{figure}
\includegraphics[width=\columnwidth]{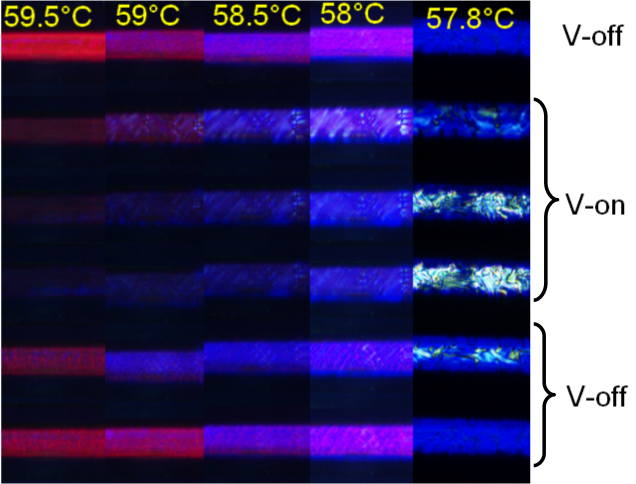}
\caption{BPLC capillary with different temperature under the influence of electric field.} 
\label{fig:EF_temp_BP}
\end{figure}

Next, the BPLC capillary with NAL was put between glass plates with ITO (Indium Tin Oxide) electrodes. In the subsequent measurements the external electric field was applied right after the stabilization of temperature, without waiting for BP domains growth. In Fig.\ \ref{fig:EF_BP} the top most image corresponds to BP\,II at $59^{\circ}\, \text{C}$ and 0\,V. To obtain electric field between ITO electrodes the high voltage amplifier and generator were used. For voltage values less than 900\,$\text{V}_\text{{pp}}$ no impact on the BP domains was observed. For 1000\,$\text{V}_\text{{pp}}$ and 1\,kHz square wave signal, the uniform BP\,II structure was obtained. The electric field (EF) inside BPLC capillary was 2.5 $\text{V}/ \mu \text{m}$. The BP capillary started to reflect red light. This process was not rapid, after 1 minute of applying the external voltage, the reorientation process of BP phase was observed (tinted images in Fig.\ \ref{fig:EF_BP}). Then, the voltage was switched off, and after 1 minute a stable, uniform structure, strongly reflecting red light was observed. This state was persisting as long as the constant temperature was maintained for BP\,II. Subsequently, the influence of the electric field on selective light reflection in BPLC capillary for different BP temperatures and phases was investigated. In Fig.\ \ref{fig:EF_temp_BP} the images of red, pink-violet and blue light reflections under the influence of the electric field are presented. 
It has been noticed that for BP\,II at $59.5^\circ \,\text{C}$ the applied voltage weakens the reflection of light. In this situation, the EF starts unwinding BP\,II structure, but the forces are so strong so the BP\,II returns to the initial state. 
Under the influence of electric field in BP\,I at $57.8^{\circ}\,\text{C}$ the chiral phase was induced. This effect was repeatable, and it was possible to switch from the BP\,I to the chiral phase. For VA and HA in BP capillary structure the same results were observed. After applying high voltage the structure also reflected red light. Only for VA and HA the changes of selective light reflection, already appear in the range from 400\,$\text{V}_\text{{pp}}$ to 600\,$\text{V}_\text{{pp}}$.  

\section{Conclusions}
For the first time, the process of BP domains growth in a capillary was observed. A quasi uniform structure in the BPLC capillary was obtained by slow cooling process. Also the uniform structure was achieved by applying the external electric field. The changes of selective light reflection by using NAL, HA, VA and the electric field were observed. The ability of switching between BP\,I and chiral phase in a capillary was also shown. The presented results can be, for instance, useful in designing a new types of optical attenuators or optical switches based on the photonic crystal fibers filled with BPLC. The crystals of BPLC in a capillary could be potentially used as an equivalent of the Bayer filter in Liquid Crystal Displays (LCDs).    
\section{Acknowledgments}
This work has been supported by the National Science Centre, Poland under the grant number 2015/19/D/ST3/02432
\newline
\bibliographystyle{elsarticle-num-names} 
\bibliography{literature}
\end{document}